\documentclass[letterpaper,12pt]{article}
\usepackage{amsmath}
\usepackage{amsfonts}
\usepackage{amssymb}
\usepackage{graphicx}
\usepackage{physics}
\usepackage{latexsym}
\usepackage{epsfig}
\usepackage{pstricks}
\usepackage{stmaryrd}
\usepackage{rotating}
\usepackage[english]{babel}
\usepackage{setspace}
\usepackage[utf8]{inputenc}
\usepackage{amsmath}
\usepackage{braket} 
\usepackage{hyperref}
\usepackage{authblk}
\usepackage{natbib}
\usepackage[T1]{fontenc}
\usepackage{lmodern}
\usepackage{enumitem}
\usepackage{csquotes}
\usepackage{amsthm}
\usepackage{xcolor}
\usepackage[normalem]{ulem}
\begin{document}
%%%%%%%%%%%%%%%%%%%%%%%%%%%%%%%%%%%%%%%%%%%%%%%%%%%%%%%%%%
%%%%%%%%%%%%%%%%%%%%%%%%%%%%%%%%%%%%%%%%%%%%%%%%%%%%%%%%%%
\begin{center}	
\begin{LARGE}
\textbf{A reply to Rovelli's response to our ``Assessing Relational Quantum Mechanics''}\\
\end{LARGE}
\end{center}

\begin{center}
\begin{large}
R. Muciño, E. Okon and D. Sudarsky\\
\end{large}
\textit{Universidad Nacional Aut\'onoma de M\'exico, Mexico City, Mexico.}\\
\end{center}

In a recent paper, Rovelli responds to our critical assessment of Relational Quantum Mechanics (RQM). His main argument is that our assessment lacks merit, because we fail to understand, or cope with, the premises of his theory; instead, he argues, we judge his proposal, blinded by the preconceptions inherent to ``our camp''. Here, we explicitly show that our assessment judges RQM on its own terms, together with the basic requirements of precision, clarity, logical soundness and empirical suitability. Under those circumstances, we prove false Rovelli's claim that RQM provides a satisfactory, realistic, non-solipsistic description of the world. Moreover, his reply serves us to further exhibit the serious problems of the RQM proposal, as well as the failures of its author to understanding the basic conceptual difficulties of quantum theory.\\

%\tableofcontents
\onehalfspacing
%%%%%%%%%%%%%%%%%%%%%%%%%%%%%%%%%%%%%%%%%%%%%%%%%%%%%%%%%%
%%%%%%%%%%%%%%%%%%%%%%%%%%%%%%%%%%%%%%%%%%%%%%%%%%%%%%%%%%
%\section*{Introduction}
%\label{Intro}
%%%%%%%%%%%%%%%%%%%%%%%%%%%%%%%%%%%%%%%%%%%%%%%%%%%%%%%%%%
%%%%%%%%%%%%%%%%%%%%%%%%%%%%%%%%%%%%%%%%%%%%%%%%%%%%%%%%%%

In [2], we present a critical assessment of Relational Quantum Mechanics (RQM). In a recent response (``A response to the Mucino-Okon-Sudarsky's Assessment of Relational Quantum Mechanics'', arXiv:2106.03205), Rovelli argues that our assessment ``presupposes assumptions that are precisely those questioned in the Relational Interpretation, thus undermining the value of the assessment.'' Moreover, he promotes as an explanation for our inability to abandon such presupposed assumptions the fact that (some of us) have worked on objective collapse models.

In this reply to Rovelli's response, we explicitly show that our assessment judges RQM on its own merits, together with proposal-independent requirements, such as precision, clarity, logical soundness and empirical suitability. Under these terms, we show that Rovelli's claim that RQM provides a satisfactory, realistic, non-solipsistic description of the world is untenable. Moreover, we use his reply to further exhibit the serious problems of the RQM proposal, as well as Rovelli's failures to understanding the basic conceptual difficulties of quantum theory.

In order to analyze Rovelli's response, and to make it absolutely clear that we are not misrepresenting his proposal---judging it, as Rovelli claims, by our ``own prejudices''---we find it convenient to transcribe in full Rovelli's response, and to reply in detail to each one of his claims immediately below his own words. We put Rovelli's words in {\bf bold face} and use normal type in our comments and replies to his claims. It is important to point out that, here, we concern ourselves with specific aspects of our assessment that were touched by Rovelli in his response; for a full evaluation, we redirect the interested reader to our comprehensive initial assessment in [2].

\section*{I. THE PROBLEM OF QUANTUM PHYSICS}
\label{I}

\hspace{\parindent} {\bf The problem of quantum physics is not that we have \emph{no} way of making sense of it. The problem is that we have \emph{many} ways of making sense of it. But each of these comes with a high conceptual price.}

{\bf Each interpretation of quantum mechanics demand us to accept conceptual steps that for many are hard to digest. Pilot-Wave like interpretations require a non-local layer of reality which is inaccessible in principle; Many-Worlds interpretations require zillions of real actual copies of ourselves seeing slightly different worlds; Qbism forces us to a strong instrumentalism; Physical Collapse models require physical processes that have never been observed; Relational Quantum Mechanics assumes that contingent properties are sparse and relative. And so on.}

Rovelli claims that we have many ways of making sense of quantum theory. We, on the other hand, would argue that, while there are many \emph{proposals} for making sense of the theory, not all of them belong in the same basket. We agree with him in that all proposals contain elements that many might find hard to digest. However, we maintain that not all proposals fare equally well on a list of basic, proposal-independent, criteria, which allow for an objective assessment of the viability of a given framework. Such proposal-independent criteria are requirements that cannot be negotiated in the construction of a sensible fundamental physical theory, and include demands for clarity and precision in the formulation of the proposal, self-consistency, empirical adequacy, or the avoidance of unreasonable scenarios such as solipsism.\footnote{One could argue that, by relinquishing a realist position---broadly speaking, the idea that there is a world out there, and that science offers a path to (at least approximately) characterize it---one eliminates the requirement to satisfy some of these criteria. That may be so, but since Rovelli is clear that RQM should be given a realist reading, RQM must be evaluated against all of these criteria.} Moreover, it is also true that none of the proposals on the market are polished, final products, which means that all of them contain open issues that need to be sorted out before declaring them truly successful. The upshot of all this is that, in order to properly explore issues in the realm of quantum foundations, it is key to differentiate between the merely unpalatable aspects of a proposal, open questions that need further analysis, and failures at the level of the basic, proposal-independent criteria.

For instance, it is widely accepted within the field of quantum foundations that the standard interpretation does not fare well against some of these criteria. Among other things, the theory is problematic due to its reliance on the vague notion of ``measurement'' (or some equally imprecise concept). Still, no one doubts that the theory is extremely successful in practice. This is because, in most practical applications, physicists have learned to evade its problems, basically by judiciously \emph{supplementing} the theory with some practical knowledge, not contained or derivable from the standard framework alone---such as the fact that we never observe macroscopic objects in superposition of different positions. Because of this, people interested in foundational aspects, readily recognize that the standard theory is not really satisfactory. Then, when one sets for oneself the goal of proposing a novel approach to deal with these problems, the evaluation of the proposal must be carried on by checking whether it really deals with them, instead of simply perpetuating the unsatisfactory approach of the standard framework. 

In sum, the problem of quantum physics is not that we have many ``equally reasonable'' ways of making sense of it. We have some well-developed proposals, with lingering open issues, but we also have other approaches, which face much more profound difficulties, at the level of the basic criteria. By allowing the presence of vague, inconsistent or solipsistic proposals in the basket, we only create more confusion and help perpetuate the alarmingly low levels of conceptual clarity prevalent in many discussions of quantum foundations.

{\bf Richard Feynman observed that Nature often admits different interpretations of the same phenomena. He suggested that a good scientists should better keep all alternative in mind, not knowing which one will next turn out to be good. Perhaps this is a good attitude for quantum theory. As science develop, one perspective may well turn out to be more fruitful. One price will turn out to be worth paying. In one form or the other, all approaches are radical (because the novelty of quantum theory to be radical), but the conceptual assumptions of the different interpretations are in general radically different. For this reason, communication between scientists and philosophers working within different interpretations is sometimes uneasy: it often reduces to an empty restatement of alternatives ``beliefs''.}

We could not agree more strongly with Feynmann's recommendation, but we are sure he omitted to state the important caveat, it being self-evident, that we should proceed in that manner, but only with views that are, well-defined, logically coherent and potentially empirically viable.

{\bf This is unfortunately the case of the ``assessment'' of Relational Quantum Mechanics (RQM) [1] recently attempted by Muciño, Okon and Sudarsky (MOS) [2]. These authors are defenders of Physical Collapse models [3–6]. They give for granted a number of conceptual assumptions that ground the Physical Collapse interpretation of quantum mechanics.}

{\bf They evaluate RQM starting from the conceptual assumptions on which their own work is based, and find that RQM does not solve \emph{their} problems.}

{\bf MOS fail to see that RQM is coherent (as other interpretations are), but at the price of giving up a priori assumptions that they are not ready to give up. Hence their ``assessment'' of RQM end up being empty: they do find contradictions, but the contradictions are not within RQM; they are contradiction between the RQM and assumptions that are given for granted by the author's, but are not commonly accepted.}

{\bf In this paper, I discuss in detail the assessment of RQM that they give, and explain in detail why it misses the point.}

In these two paragraphs, we see the core of Rovelli's defense: he accuses us of being unable to produce an objective assessment of RQM because, by (a couple of us) having worked on objective collapse models, he believes we are unable to leave behind the conceptual assumptions behind such a program. As a result, Rovelli claims, we do not find RQM to be unsatisfactory, we simply find it unpalatable because it is not an objective collapse model.

The first thing we point out is that, as we show bellow in detail, Rovelli is never able to exactly point out what are these conceptual assumptions he believes are behind our assessment. He just talks vaguely about some sort of fixation we have with objective collapses, but is unable to point to a specific use of such assumptions in any of our objections against RQM. Moreover, there is a quick and straightforward way to see that Rovelli's defense simply cannot be right. As he correctly points out in his reply, in our paper we use pilot-wave theory, a framework with exact unitary evolution at all times, and no collapses anywhere, as an example of a satisfactory proposal. That would be really strange if, as Rovelli claims, we are indeed engrossed with objective collapses. The truth, as we show bellow in detail, is that we do judge RQM on its own terms and merits, and we simply find it wanting.

Before proceeding, we clarify that, although in these paragraphs Rovelli talks about us finding inconsistencies within RQM, we actually never claim to do so. What we argue, by analyzing a simple, symmetrical EPR scenario, is that RQM is either incomplete or solipsistic (bellow we come back to this issue in detail).

\section*{II. RQM IN A NUTSHELL}

\hspace{\parindent} {\bf This is not the place for a detailed description of RQM, for which I refer to the literature [1, 7–9]. But for the purpose of clarity, I give here a compact account of this interpretation (as I understand it today).}

{\bf The basic idea is that the world can be decomposed (in many alternative manners) into ``physical systems'' that interact among themselves. Each physical system can be characterized by a set (in fact, an algebra) of physical variables $A_1,..., A_n$. These variables do not represent how the system \emph{is}, but rather what the system \emph{does} to another system, when there is an interaction. Outside this context, they are not determined [10, 11].}

{\bf Physical variables (i) take value only at interactions and (ii) the value they take is relative to the interacting system. The occurrence that a variable takes a value at an interaction is called a quantum event.}

{\bf For a given system $S$, its quantum theory specifies its variables $A_n$, the set of values that each of these can take (that is, their spectrum), and gives probabilities $P(a_n; b_n)$ for values $a_n$ \emph{relative to a second system} $O$, as a function of values $b_n$ relative to this same system $O$.}

{\bf Values taken by variables are labelled by the systems in interaction, but it can be shown [9] that the effect of decoherence is to render the labelling irrelevant as soon as quantum interference is suppressed.}

{\bf Key features of this interpretation, important for what follows are:}

\noindent {\bf (a) The interpretation is realist in the sense that it describes the world as a collection of real systems interacting via discrete relative quantum events.}

We must say that we find it hard to make sense of this statement. Interactions, even within RQM, are described in terms of suitable Hamiltonians or Lagrangians, while events, in the sense considered by Rovelli, are the actualizations of values during interactions. Probably, what Rovelli meant, is that the interpretation is realist in the sense that it describes the world as a collection of real systems interacting, giving rise to discrete relative quantum events. Those events, by the way, have no mathematical description in the theory; we are just told that they are what accompanies actualization of values during interactions. Therefore, the only mathematical structure available in RQM in order to characterize them is to write something like $ \alpha \ket {1} + \beta \ket {s} \to \ket {1}$, i.e., just what is usually described as a collapse but, this time, applied to relative states.

{\bf (b) The wave function is interpreted epistemically, in the same manner in which its classical limit, namely the Hamilton-Jacobi function, is.}

Once more, this statement is hard to understand, among other reasons, because the Hamilton-Jacobi function in classical mechanics is usually argued to play a \textit{nomological} role, prescribing the physical evolution of the system, and not an epistemic one. We wonder whether Rovelli might be confusing these two notions, and what he means is that the wave function plays a nomological role. Otherwise, he might be in trouble, because, it seems quite clear that a proton, with respect to which an electron might have a relative wave function according to RQM, cannot sensibly be said to ``know'' something about the electron.

{\bf (c) There are no special systems playing the role of ``observers'', no special role given to ``agents'', or ``subjects of knowledge'', no fundamental role given to special ``measurement'' contexts.}

This is fine and even desirable in any proposal that attempts to address the shortcomings of the standard interpretation. We, however, cannot fail to note that this statement is simply in contradiction with the former, if the word ``epistemic'' is given its usual meaning---a fact that might confirm our claim (assuming Rovelli does not want to contradict himself so directly in two consecutive statements) that he might be using \emph{epistemic} incorrectly.

{\bf (d) The traditional tension between unitary evolution and wave function collapse is resolved by relativising values. That is, the evolution of the probability distribution of the values of variables \emph{relative to a system} $O$ is predicted by unitary evolution only as long as $S$ is isolated and does not interact with $O$. This does not conflict with the fact that variables of subsystems of $S$ can take value with respect to one another, because the quantum mechanical transition amplitudes only connect values relative to the same system. In a slogan: with respect to Schr\"odinger's cat the poison is definitely out or not, but this has no bearing on the possibility of an external observer to observe quantum interference effects between the two alternatives.}

As we show in [2], RQM's ``resolution'' of the traditional tension between unitary evolution and wave function collapse is as obscure and vague as in standard quantum mechanics. The tension is dressed with new words such as ``any interaction leading to an event'' instead of ``measurement'' but the problem prevails. Even worse, Rovelli's proposal leads to serious conceptual problems of its own, such as the fact that it cannot even explain experimental results because, as we show below in detail, the comparison of descriptions between different observers becomes either unexplained or senseless.

{\bf (e) A coherent picture of the world is provided by all the values of variables with respect to \emph{any single} system; juxtaposing values relative to \emph{different} systems generates apparent incongruences, which are harmless because they refer to a non-existing ``view from outside the world''. These minimal notes should be sufficient to address the points raised by MOS.}
 
We are certainly willing to contemplate the idea that there is no ``view from outside the world''. However, we do require for different observers to be able to compare their different descriptions in a sensible manner, not just for there to be no contradictions for self-descriptions. If such comparisons cannot be explained with the structures provided by the theory, we are forced to conclude the theory is either incomplete (i.e further structure is needed to ensure no contradictions arise) or is solipsistic (because different agents will have disconnected accounts of what actually takes place in the world). 

\section*{III. HOW MOS DEFINE THE PROBLEM OF QUANTUM MECHANICS}

\hspace{\parindent} {\bf The characterization of the problem of quantum mechanics (QM) that MOS give is idiosyncratic. It makes sense if we remember that they expect that the only way to solve it is to introduce a physical collapse mechanism. According to MOS, the problem of QM is indeed the following: say a Stern-Gerlach apparatus entangles the position of a particle with the $z$ component of the spin. Schematically, with obvious notation:}
\begin{equation}
	\ket{\psi} \sim \ket{1}\otimes \ket{+}_z + \ket{2}\otimes\ket{-}_z
\end{equation} 
{\bf This same state can also be written as}
\begin{equation}
	 \ket{\psi} \sim \left(\ket{1} + \ket{2} \right) \otimes \ket{+}_x + \left(\ket{1} + \ket{2} \right) \otimes \ket{-}_x 
\end{equation}
{\bf The problem of QM, according to MOS is: why then this same apparatus does not measure the $x$ component of the spin? In their words: ``Without information not contained in a complete quantum description, standard quantum mechanics is unable to deliver concrete predictions regarding the possible final outcomes of an experiment.''}

{\bf This is a bizzarre characterization of the problem. Of course standard quantum mechanics is able to deliver concrete predictions regarding the possible final outcomes of an experiment, as any experimenter can testify. The reason is that ``standard quantum mechanics'' is not only a quantum state evolving unitarily. Standard quantum mechanics as formulated in textbooks (mostly in the Copenhagen language) includes much more: operators associated to measuring procedures, eigenvalues, the existence Bohr's macroscopic world and so on. Once you fold these ingredients in, predictions are clear.}

There is a lot to unpack here. To begin with, Rovelli's description of our take on ``the problem of quantum mechanics'' is simply laughable. We devote several pages to carefully explain what we take to be the different conceptual problems of the standard interpretation, but Rovelli takes one aspect of one of the issues, mischaracterizes it, and presents it as our whole take on the issue. Moreover, he argues that our characterization of the problem only makes sense from the perspective of objective collapse models but he never really explains why he thinks that to be the case. Regardless, it is easy to see that this cannot be true since, as we said above, we take theories without collapses, such as pilot-wave, as perfectly valid solutions to the problems we describe.

Now, regarding the aspect that he takes as our characterization of the problem, he claims that when we talk about ``a complete quantum description'' we are only talking about a quantum state evolving unitarily. Such a claim is either completely careless or misleading, as we clearly explain in our assessment that, by a complete quantum description, we mean ``a full description of the quantum state of the measuring apparatus and all the details of its interaction with the rest of the system''. Our point, of course, is that even if one is given ``operators associated to measuring procedures, eigenvalues, the existence Bohr's macroscopic world and so on'' the standard framework is unable to deliver concrete predictions regarding the possible final outcomes of an experiment. As we explained above, in practice, physicists do make concrete predictions but, to do so, they (often inadvertently) employ practical information that is not contained within the actual theory.

In the original assessment, we provide a simple example that, apparently, Rovelli did not understand. Here we try again with a different one. Actually, the example we present next was presented by one of us personally to Rovelli at a meeting,\footnote{Private communication at the meeting ``Do Black Holes exist? The physics and philosophy of Black Holes'', Physikzentrum, Ban-Honnef, Germany, April 2017.} so it has the advantage that we know for sure that he clearly understood the issue. We frame the problem in the usual language for the reader's convenience, but we make sure to reformulate it in the way RQM requires.

The simple example is the following. Take a free particle with Hamiltonian
\begin{equation}
		\hat{H}_p^{0} = \frac{\hat{p}^2}{2m}
	\end{equation}
which, at $ t=0$, is in the state $\psi (x,0)$, given by a sharply peaked wave packet centered at $x=0$ (and symmetric under $ x\to-x$). [This particle is latter going to play the role of system $ S$ according to RQM, and the state will be taken as relative to $O$, a detector system to be specified]. Next, consider two 2-level detectors (with basis $|-\rangle$ (ground) and $|+\rangle$ (excited)), located at $x=D$ and $ x = -D$. [The detector pair will play the role of $ O$ in the RQM language]. Assume that, at $ t=0$, both detectors are in the ground state. The free Hamiltonian of each detector is
	\begin{equation}
		\hat H_i^{0}= \epsilon \lbrace | + \rangle {}^{(i)} \langle + | {}^{(i)}- | - \rangle {}^{(i)} \langle - |{}^{(i)} \rbrace.
		%\right],
	\end{equation}
where $i=1, 2$. Finally, the Hamiltonian describing the interaction of particle and detector $i $ is 
	\begin{equation}
		\hat H_{i} ^{I} = \frac{g}{\sqrt {2}} 
		\delta (x\pm D \hat{I_p}) \otimes (| + \rangle {}^{(i)}\langle - |{}^{(i)}
		+ | -\rangle{}^{(i)} \langle + |{}^{(i)}) 
	\end{equation}
Note that the initial conditions and the total Hamiltonian are symmetric under $x \to -x$.

Summing up, the initial state of the $S-O$ system, relative to a third system (say, a nearby particle, playing the role of system $P$) is given by
\begin{equation}
	\Psi (0) = | \psi_0 \rangle \otimes | - \rangle {}^{(1)}\otimes | - \rangle {}^{(2)} 
\end{equation}
and it is easy to see that, after some time $t$, the state will evolve to
\begin{eqnarray}
	\Psi (t) & = &
	| \psi_{+-} \rangle\otimes | + \rangle {}^{(1)}\otimes | - \rangle {}^{(2)} +| \psi_{-+} \rangle\otimes | - \rangle {}^{(1)}\otimes | + \rangle {}^{(2)} \nonumber \\
	& & + | \psi_{--} \rangle \otimes | - \rangle {}^{(1)}\otimes | - \rangle {}^{(2)} + | \psi_{++} \rangle \otimes | + \rangle {}^{(1)}\otimes | + \rangle {}^{(2)} ,
\end{eqnarray}
which, of course, preserves the symmetry $x \to -x$. It would seem that one can interpret each of the terms above easily, and make the corresponding predictions: either detector 1 was excited, detector 2 was excited, no detection has taken place (because no detector can be 100\% efficient) or there has been a double detection (involving a bounce and characterized by a small number $ O(g^2)$). However, the point we want to make, is that one might consider, instead, describing things in terms of an alternative basis for the detectors pair given by
\begin{equation}
	| U \rangle \equiv | + \rangle^{(1)} \otimes | + \rangle^{(2)} 
\end{equation}
\begin{equation}
	| D \rangle \equiv | - \rangle^{(1)} \otimes | - \rangle^{(2)} 
\end{equation}
\begin{equation}
	| S \rangle \equiv \frac{1}{\sqrt 2 } [ | + \rangle^{(1)} \otimes | - \rangle^{(2)} + | - \rangle^{(1)} \otimes | + \rangle^{(2)}]
\end{equation}
\begin{equation}
	| A \rangle \equiv | \frac{1}{\sqrt 2 } [ | + \rangle^{(1)} \otimes | - \rangle^{(2)} - | - \rangle^{(1)} \otimes | + \rangle^{(2)}] 
\end{equation}
Moreover, the Hamiltonians can be expressed in terms of this basis. For instance, the full interaction Hamiltonian 
\begin{eqnarray}
\label{Int-Ham1}
	\hat H^{I} & = & \frac{g}{\sqrt {2}} \lbrace ( \delta (x - D \hat{I_p}) \otimes (| + \rangle {}^{(1)}\langle - |{}^{(1)}
	+ | -\rangle{}^{(1)} \langle + |{}^{(1)}) \otimes I^{(2)} \nonumber \\
	& & \qquad + \delta (x + D \hat{I_p}) \otimes I^{(1)}\otimes (| + \rangle {}^{(2)}\langle - |{}^{(2)} + | -\rangle{}^{(2)} \langle + |{}^{(2)}) \rbrace
\end{eqnarray}
can easily be rewritten as
\begin{eqnarray}
\label{Int-Ham2}
	\hat H^{I} & = & \frac{g}{2} \lbrace (\delta (x - D \hat{I_p}) +\delta (x +D \hat{I_p}) )\otimes [ | U \rangle \langle S | +
	| S \rangle \langle U | + | S \rangle \langle D | + | D \rangle \langle S | ] \nonumber \\
	& & + (\delta (x - D \hat{I_p}) -\delta (x +D \hat{I_p}) )\otimes [ | A\rangle \langle D | + | D \rangle \langle A | -
	| U \rangle \langle A | - | A \rangle \langle U | ] \rbrace
\end{eqnarray}
It is clear, then, that the Hamiltonian can be written in both sets of variables. Therefore, if the Hamiltonian is what describes the interaction, then it is of no help in selecting among the possible choices (in clear contradiction of Rovelli's claims above).

In sum, it is clear that, without further input, beyond that encoded in the full quantum description, including the state of the system, the free and interaction interaction Hamiltonians, all operators, eigenvalues, etc., there is nothing that would allow us to decide what description is the appropriate one: the one that uses for the detector pair the basis $\lbrace | + \rangle^{(1)} \otimes | + \rangle^{(2)} ,\ | - \rangle^{(1)} \otimes | - \rangle^{(2)} ,\ | + \rangle^{(1)} \otimes | - \rangle^{(2)}, \ | - \rangle^{(1)} \otimes | + \rangle^{(2)} \rbrace $ or that which uses the basis $\lbrace | U \rangle,\ | D \rangle,\ | S \rangle ,\ | A \rangle \rbrace $).

Of course, if we allow into the discussion additional elements, such as intuition or practical experience, we would have no problem deciding that the appropriate choice is provided by the first description. The problem lies in the fact that, such additional elements, are not part of the fundamental theory, be it the standard framework or RQM. Then, there must be more to physics than what can be encoded in states (relative or not) and the Hamiltonians characterizing their dynamics and interactions. Thus, we must conclude that such frameworks are (at best) incomplete. Moreover, it is clear that this issue has nothing to do with collapse theories, or with what Rovelli calls ``our camp''. It is part of the measurement problem, a problem that, apparently, Rovelli does not fully understand. As for \emph{solutions} of such a problem, the challenge is for the theory to be able, among other things, to \emph{predict} the correct basis out of a precisely formulated theory, without the need of our practical knowledge of which macroscopic superpositions are never experienced in the lab. That is, in the example above, the theory would have to explain and justify which one of the multiple options, including the two explicitly presented above, is the appropriate one.

{\bf Of course there are good reason to consider the Copenhagen interpretation unsatisfactory, and interpretations that have a larger applicability arguably exist. With the possible exception of Many Words, in general other interpretations do \emph{not} start by simply discarding everything except the evolving state. Hence the characterization of the problem given by MOS is, at most, a potential objection to Many Worlds, certainly not the general formulation of the problem of quantum theory.}

This comment fully depends on the mischaracterization by Rovelli of our description of the problem. However, as we just saw, there is a real problem (that Rovelli simply appears not to understand) which is, both a challenge to standard quantum mechanics in general, and one that RQM cannot deal with. The issue is not just that of looking for a theory that might have ``larger applicability'', but that the theory, as it stands, cannot be regarded as fundamental or complete.

{\bf What equations (1) and (2) show is that there are (special) cases in which the bi-orthogonal decomposition of a state is not unique. This is well known. It is a potential problem only for the interpretations of quantum mechanics that rely on the bi-orthogonal decomposition, such as modal interpretations [12–14] or some versions of Many World. The bi-orthogonal decomposition theorem plays no role in many other interpretations and in particular in RQM, which is about variables, not about states.}

It should be clear by now that, contrary to what Rovelli claims above, the problem under discussion has absolutely nothing to do with bi-orthogonal decomposition.

{\bf More explicitly, this is the definition of the problem of QM given my MOS: ``The formalism [of RQM] ends up critically depending upon the notion of measurement---which is a problem because such a notion is never precisely defined within the theory. And it is not only that the standard theory does not specify when a measurement happens, it also does not prescribe what it is that is being measured (i.e., in which basis will the collapse occur).''}

{\bf Notice that the problem is \emph{formulated} in terms of ``measurement''. But the notion of ``measurement'' plays no role in RQM. In fact, the entire logic of RQM is to give up any notion of ``special'' interactions that should count as ``measurements''.}

Here, once more, Rovelli is being either completely careless or straight-out misleading. The problem is that he takes what, in our assessment, is unmistakably described as a problem with the \emph{standard interpretation}---which, of course, does depend upon the notion of measurement---but he, with full liberty, inserts ``[of RQM]'' to change our words and make it appear that we are attributing such a problem to his proposal. At any rate, we are fully aware that the notion of measurement is argued to play no role in RQM. The problem, as we clearly explain in our assessment, is that Rovelli's attempt to substitute ``measurements'' by ``interactions'' does not really solve the ambiguity issue because, at the end, RQM does not provide enough resources to determine what counts as an interaction, when interactions happen, and which variables acquire values.

{\bf Since MOS work in Collapse Models, their assumption is that that reality is described objectively and universally by an evolving wave function. This wave function evolves unitarily until something ``special'' happen. This special is the ``measurement''. Recall that in Collapse Models something (a fundamental frequency [15], a threshold in the gravitational self-potential [16],...) must determine the occurrence of the physical collapse. Under this logic, MOS asks whether there is something ``special'' that determines when a measurements happen in RQM. The answer is that there is no ``measurement'' in RQM. There is no universal objective wave function either, in RQM.}

Of course, our point here has absolutely nothing to do with collapse models. Remaining strictly within RQM, Rovelli talks about relative actualizations of values, i.e., quantum events, and the issue is that such quantum events are simply ill-defined in the absence of a way to determine, albeit probabilistically, when they happen. The issue, once more, is that Rovelli's recipe for doing so does not work---unless it is supplemented with information not contained within what RQM takes to be a full description.

{\bf The formulation of the problem of QM according to MOS is predicated on the basis of assumptions that are explicitly rejected in RQM.}

This is simply nonsense. As we saw above, to arrive at this conclusion, Rovelli cherry-picks, twists and misrepresents our words and arguments. The truth is that, in our assessment, we explicitly acknowledge that there is no universal and objective wave function, and our objections at no point rely on ignoring such a fact. We analyze various aspects of the proposal, accepting the validity of any description (in terms of relative states) made by any observer (which, according to RQM, can be any system at all) and consider in detail the story Rovelli tells about the acquisition of relative values. The problem we find is that Rovelli's story is simply inadequate; as inadequate as the collapse of the wave function in textbook accounts.

Moreover, as we already explained, this problem of RQM has nothing to do with collapse models or universal wave functions. There are, in fact, satisfactory proposals for making sense of quantum physics, such as pilot-wave, that assume that wave functions always evolve unitarily, but still have a clear explanation of how observers find definite values for certain variables. No collapse needed here. One would expect the same amount of clarity in RQM, a theory that claims to be a viable and complete alternative. We do not demand for RQM to have a universal wave function or an objective collapse, but to offer a reasonable response to the traditional difficulties of quantum theory.

\section*{IV. WHICH VARIABLE TAKES VALUE}

\hspace{\parindent} {\bf Nevertheless, we can still try to translate the question posed by MOS into the language of RQM and see if it may refers to anything. The question could be which variable of a system takes value relative to another system (that is, when does a quantum event happen), and when does it do so.}

{\bf Recall that the setting of RQM is not a uniformly evolving quantum state. It is a setting in which two specific distinguishable physical systems are singled out, say S and P. Quantum mechanics gives descriptions of the world conditional to this (arbitrary) choice and describes how one system affects the other \emph{when they interact}. The theory can therefore be applied anytime we have two well distinct systems interacting.}

{\bf Which variable takes value in the interaction is dictated by the physics: in the classical theory, we can describe the interaction between the two systems, say, in terms of an interaction term in the Hamiltonian that depends, in particular, say, on a variable A of the system S: then A is the value that takes value. The reason is that the interaction Hamiltonian depends on the property of S responsible in determining the effect of S on O. And this is precisely how quantum theory describes the world (in RQM): the way systems affects one another.}

{\bf The formulation of the problem in terms of the states (1) and (2) above does not even make sense in RQM: the theory is about values of variables, not about states.}

We can use the example described above to easily show that things do not work as Rovelli claims. The example contains two distinct systems interacting, the particle, which we call $ S $, and the pair of detectors, which we call $O$. We have the state of the $S-O$ system, which we take to be the relative state with respect to some other system $P$, that might, according to Rovelli be anything or anyone. We are given the interaction Hamiltonian explicitly, which can, as we showed, be written in various basis; for instance, as i) Eq. (\ref{Int-Ham1}), or as ii) Eq. (\ref{Int-Ham2} ).

Now, Rovelli tells us that the variable $A$ that appears in the Hamiltonian is the one that takes values but, as we see, there is an evident ambiguity in identifying $A$, given the Hamiltonian. We would be tempted, if looking at (i), to say that the variable $A$ corresponds to the projection operators
$ | + \rangle^{(1)}   | + \rangle^{(2)} {}^{(1)}\langle + |   {}^{(2)}\langle + | $, 
$| + \rangle^{(1)}   | - \rangle^{(2)} {}^{(1)}\langle + |   {}^{(2)}\langle - | $, 
$| - \rangle^{(1)}   | + \rangle^{(2)} {}^{(1)}\langle - |   {}^{(2)}\langle + | $ and 
$| - \rangle^{(1)}   | - \rangle^{(2)} {}^{(1)}\langle - |   {}^{(2)}\langle - | $. However, while by looking at (ii), we would say that the variable $A$ corresponds to the operators 
$| U \rangle \langle U |$ , $ | D \rangle \langle D |$, $| S\rangle \langle S |$, and $| A \rangle \langle A | $. Of course, it is true that experimental colleagues would tell us that the appropriate choice is that of (i), but in doing so they would be using their intuition and experience, both of which are \emph{not} encoded in the theory. That is, by relying just on the elements explicitly used in characterizing the theory, we have no means of identifying the variable that plays the role of $ A$, as Rovelli recommends. The need to rely on external input indicates, at best, that the theory is incomplete.

In sum, the interaction Hamiltonian can be expressed in either set of variables, so Rovelli's claim that the physics, as encoded in the interaction Hamiltonian, resolves the issue, is simply false. Alternatively, we might conclude that, according to Rovelli, there is more to physics than what is encoded in the Hamiltonian and the relative states, and thus, that RQM is incomplete.

\section*{V. WHEN DOES A VARIABLE TAKE VALUE}

\hspace{\parindent} {\bf To the question of the \emph{time} when the quantum event happens, the answer is similar. It happens when the systems interact. In turn, we may ask when do the systems interact. The answer is the (quantum) physics of $S$, given its dynamics and the interaction terms in its Hamiltonian. For instance, suppose that $S$ is an electron in a radioactive atom and $P$ is a Geiger counter. When does a quantum event relative to these two systems happen?}

{\bf Notice that the Schrodinger wave function of the electron leaks \emph{continuously} out of the atom and is therefore constantly in causal contact with the Geiger counter. The electric force of the electron on the Geiger counter, irrespectively of the position of the electron is never exactly zero. If you start with a wave function ontology you have the problem of understanding what happens when the Geiger counter clicks and when it does so. But the RQM ontology is not the continuous wave function leakage. It is the actual quantum event of the clicking of the Geiger counter, which is discrete. When does it happen? Knowing the Hamiltonian of the system and the interaction Hamiltonian with the affected system, standard quantum mechanics can be used to determine the probability distribution in time of the occurrence of the interaction. That is, quantum events are discrete, their occurrence can be predicted only probabilistically, and the probability distribution of their occurrence can be computed using quantum mechanics itself and the specific quantum dynamics in play.}

{\bf The difficult conceptual step here is to accept the idea that continuity is a large scale approximation, while the happenings of the world are discrete and probabilistic at the quantum scale. In its relational interpretation, quantum theory describes a world that is fundamentally discrete and probabilistic. Exchanges between system are always discrete and regulated by the quantum of action $\hbar$.}

{\bf This is precisely the original intuition of Max Born and its collaborators in Gottigen [17–20], who were the first creators of quantum theory. But it is also the conclusion to which their opponent, one of the most strenuous defenders of continuity and deepest thinkers about the problems of the theory, Erwin Schrodinger, ended up accepting: ``There was a moment when the creators of wave mechanics (that is, himself) harboured the illusion of having eliminated discontinuity from quantum theory. But the discontinuities eliminated from the equations of the theory reappear the moment the theory is confronted with what we observe. [...] it is better to consider a particle not as a permanent entity but rather as an instantaneous event. Sometimes these events form chains that give the illusion of being permanent, but only in particular circumstances.''[21]}

{\bf MOS reject this logic a priori, because they interpret quantum measurements as ``special events'' that happen to a realistically interpreted wave function under peculiar circumstances to be specified, not as the general happening of all phenomena.}

{\bf Within MOS's Physical Collapse logic, RQM does not answer MOS's question. But this is not because of an incoherence of RQM, it is only because MOS have a priori assumptions that are rejected in RQM. The assumptions of RQM are similar to the original ones of Born and collaborators. The assumptions of MOS are the early ones (lately rejected) by Erwin Schr\"odinger.}

Once again, Rovelli does not engage with or refute our arguments. Instead, he rambles around these broad and vague accusations of them being dependent on our ``Physical Collapse logic''. In any case, once more, it is easy to see, with the example presented above, that his acrobatics are useless. In the example, we have the Hamiltonian of the system $S-O$, and we have the relative state of that system with respect to $P$. Can we answer when does the so called ``quantum event'' happen, or calculate the probability for it to happen in a given interval $(t, t+\delta t)$? Again, with the elements provided by RQM, plus the specific situation at hand, we have no answer to these questions. Rovelli's claim that ``the physics will determine it'' is simply, either false, or means that RQM is incomplete, i.e., that the physics includes more than what is encoded in the full description, according to RQM. As in the discussion of the previous sections, it is in fact true that most physicists would be able to say something about the issue, but what could be said is, on the one hand, rather vague and, more importantly, it would be based on intuition and experience, and not on elements codified in the theory.

\section{VI. WHEN DOES MEASUREMENT A HAPPEN?}

\hspace{\parindent} {\bf In their quest for finding a rule for an objective and system independent condition for quantum measurement, MOS make reference to an old article with a latin title ``Incertus tempus, incertisque loci: when does a measurement happen'' [22]. What MOS search, namely an objective criterium singling out quantum measurement, is not in that paper, because the paper addresses a different problem.}

{\bf In RQM terms, that paper asks the following question. Suppose that two systems $S$ and $O$ interact among themselves. If we consider \emph{only} values of variables relative to a third system,$ P$ , is there anything we can say \emph{in terms of these values} about the timing of a quantum event realised by the interactions between $S$ and $O$?}

{\bf At first sights the answer seems to be negative, because the values of the variables with respect to $P$ are blind to the values of the variables of $S$ with respect to $O$. Yet, [22] points out that there is an operational sense that can be given to this question. The intuition is that if Wigner has a friend making a measurement in a closed box, Wigner --with sufficient knowledge of what is in the box-- can say something about \emph{when} his friend makes the measurement, even if he has no access to it. This intuition is made concrete in [22] as follows. The previous interactions between the $S \cup O$ system and $P$ give probabilistic predictions about what would $P$ see (that is, how $ S \cup O$ would affect $P$ ) if an interaction with $P$ happened at some arbitrary time. Now consider an interaction where $O$'s ``pointer'' variable affects (``is seen by'') $P$. The predictions can include the timing when the pointer move, even if they do not include which direction it will moves. Hence $P$ can predict ``when the measurement happen'' under a special definition: with respect to $P$ , a measurement between $O$ and $S$ happens when $P$ can predict that the the pointer variable has moved. Since the prediction is probabilistic, this gives only a probability distribution of course. That is, figuratively, ``I say that the measurement has happened with probability $p$ if I predict that I will find the pointer moved (correlated with the $S$'s variable, which $P$ can independently detect) with probability $p$''. This peculiar operational definition does in fact correspond to what one would concretely say in a laboratory, but has nothing to do with the foundation of RQM. MOS misinterpret it as foundational.}

{\bf MOS observe that this operational definition is based on the notion of quantum event itself. This is correct, or course. Their objection is that this implies a regression at infinity. This is wrong, because the notion of quantum event does not require such indirect operational definition. The observation counts as an objection only if one asks, as MOS do, an absolute non-relational determination of quantum events, and searches it in [22]. But a non-relational definition of quantum events is exactly what RQM mechanics demands not to ask, in order to make sense of quantum physics.}

Rovelli here claims that he never argued that his paper with a latin title could be used to address the question of the timing of the events. This, however, is in \emph{direct contradiction} with his own words elsewhere. For instance, in the entry on RQM of the Stanford Encyclopedia of Philosophy, which is the first reference on RQM he offers in his response, he writes
\begin{quotation}
The flash ontology of RQM seems to raise a difficulty: what determines the timing for the events to happen? The problem is the difficulty of establishing a specific moment when say a measurement happens. The question is addressed in [my paper with a latin title], observing that quantum mechanics itself does give a (probabilistic) prediction on when a measurement happens. [1]
\end{quotation}

Of course, if RQM says the world to be made of relational quantum events, which appear when interactions between systems take place, one would expect RQM to determine clearly how and when these (relative) quantum events occur. In our assessment, we show that the paper with the latin title does not help at all with the issue. As a result, in his response, Rovelli seems to disown the claim that such a paper could help, but offers no alternative to deal with the issue (a problem which, as is clear in the quote above, he does think is important). It goes without saying that, without resources to determine when ``quantum events'' happen, RQM is left with no foundations for its own ontology. 

To conclude this section, it is important to point out that, contrary to what Rovelli claims, we never argue that the fact that he offers an operational definition for the timing of the events implies a regression at infinity. As everywhere else in his reply, he either refuses to engage with our arguments, or he constructs a straw-man of them, which he destroys with great pomp.

\section{VII. OBSERVER DEPENDENCE}

\hspace{\parindent} {\bf An explicit criticisms of MOS regards a statement contained in the paper that has inspired RQM, the 1995 paper [7]. The paper says that ``the experimental evidence at the basis of quantum mechanics forces us to accept that distinct observers give different descriptions of the same events''. MOS object that this not true and cite pilot-wave theory as contrary evidence. Here MOS are confusing two different facts. One is the assumption realised in the pilot-wave theory that there exist a universal objective state of affairs. A different one is the account that a real observer can give of a set of events. Since the pilot-wave theory is a hidden variable theory, observers do not have access to the universal objective state of affairs. This is hidden. In particular, the global wave function postulated by the pilot wave theory is in principle inaccessible to observers (otherwise the pilot wave theory could be used to make non-probabilistic predictions and beat standard quantum theory, which is not the case). In fact, what happens in the pilot wave theory is that observers can make predictions using ``effective'' quantum states. These are relative states and observer dependent. Thus confirming the observation in [7]. Therefore the MOS objection has no ground.}

Here, Rovelli is confusing two different facts. One is that, in pilot-wave theory, as in any theory, two different observers can give different descriptions of the same sequence of events, simply because one of them can be wrong. The second one is that, in pilot-wave theory, there is just one true, correct sequence of events and, whatever alternative description is given, is simply wrong. Therefore, again, we can conclude that Rovelli's claim is false: experimental evidence does not forces us to accept that distinct observers give different descriptions of the same sequence of events. (By the way, we are sure that he meant that, in pilot-wave, observers do not have access to the positions of the particle, not to the global wave-function.)

{\bf The other argument that MOS cite as evidence against the statement above is---not surprisingly---physical collapse model. It is of course true that collapse model do violate the observation in [7], but they violate the predictions of standard quantum theory as well (which is obvious by the fact that they are empirically distinguishable from standard quantum theory), while [7] states clearly that everything it says is within the validity of standard quantum theory. RQM makes sense of quantum theory (the most successful physical theory ever) as it is, not under the assumption that it fails. This can be seen directly: the detailed argument used in [7] to derive the above statement is based on the existence of certain quantum interference effects predicted by quantum theory and not predicted by Physical Collapse models.}

It is true that objective collapse models yield different predictions than the standard framework. However, it is also true that objective collapse models are compatible with all existing experimental evidence. Then, we simply fail to see how, the fact that they do not accept that distinct observers give different descriptions of the same events is a refutation of the claim that ``the experimental evidence at the basis of quantum mechanics forces us to accept that distinct observers give different descriptions of the same events'' In any case, pilot-wave does yield the same predictions than the standard framework, so we already know that Rovelli's claim is false. 

{\bf In the course of this discussion, MOS reiterate their prejudice about quantum collapse (''In sum, a core conceptual problem of standard quantum mechanics has to do with an ambiguity regarding \emph{the dynamics}.''), they interpret observer dependence as due to the fact that quantum collapse is disregarded (``Such an ambiguity allows for different observers to give different descriptions of the same events.'') and essentially blame RQM only for not adhering to their view.}

\section*{VIII. SELF MEASUREMENT}

\hspace{\parindent} {\bf A long section of the paper is devoted by MOS to criticize an observation about an assumed impossibility of self measurement in [7]. The observations about self measurement in that old paper are indeed vague. In fact these vague considerations have been abandoned in later presentations of RQM and replaced by sharper definitions [1]. For instance, in the condensed summary given above, variables are interpreted as describing the way a system affect \emph{other} systems. The point is of course that the ``R'' in RQM stands for ``relations'': RQM is an account of quantum physics in terms of relations, namely relative variables. It assumes that physics is about relative variables, describing how systems manifest themselves to other systems.}

{\bf MOS clearly misunderstand this. For instance they write ``Returning to Rovelli's statement that there is no meaning in being correlated with oneself, we must say that we find such a claim quite odd. [...] it is clear that the different parts of an observer certainly are correlated between them: her left hand is never more that 2 meters away from the right one.'' This objections betrays the misunderstanding that MOS have of RQM. In the case considered, of course one hand has a position with respect to another hand. The two hands can interact, exchange light signals, etcetera. But it is so precisely because they are two! What is the position of a single hand if there is noting else with respect to which can be defined? Position is in fact a quintessential example of a relational property in physics: it is always defined with respect to something else.}

Here, Rovelli is somehow restricting what one might call a ``system''. He has been explicit in declaring that a system might be anything, and has never required such system to have no parts, to be elementary, or to have no internal variables. Our example makes use of that freedom, explicitly allowed by RQM, and regards the person in question as the system (with the position of the hands as internal variables of that system). If one is allowed to change the rules of the game as one goes along, one can easily ``defeat'' any ``opponent''. In any event, we reiterate that, given any variable of a system, that variable is always correlated with itself, as can be illustrated by the fact that, if at any instance, we know its value, it is tautological that we do know its value. In short, Rovelli here is playing linguistic games, and being deceitful at that.

\section{IX. DECOHERENCE AND UNITARITY}

\hspace{\parindent} {\bf Another misunderstanding is in MOS's reading of the paper [9] which explore the role of decoherence in RQM. MOS write ``More recently Rovelli [...] has argued that decoherence plays a crucial role in explaining the breakdown of unitarity''. This is a misunderstanding of the paper. Decoherence plays no role in the breaking of unitarity. The unitary evolution of the probability distribution the values of the variables of the system $S$ with respect to a system $P$ holds only as long as $S$ can be considered isolated. At every interaction of $S$ with something else, unitary evolution of $S$ alone breaks down simply because $S$ is not anymore isolated. If $S$ interacts with a second system $O$, the unitary evolution of $S \cup O$ (relative to $P$ ) stills hold, but when $S$ interacts with $P$ there is definitely a breaking of the unitary evolution of its variables \emph{with respect to} $P$. Unitary evolution holds for \emph{any} system, but only as long as it refers to probabilites of interactions with something external.}

{\bf All this has nothing to do with decoherence. The role of decoherence in RQM is completely different. How is it possible that a stable world described by variables that are \emph{not} relative emerge form the quantum world where variables are always relative? The paper [9] analyses this question and shows that decoherence, which is a concrete physical phenomenon, perfectly accounts for this.}

{\bf MOS argue in detail (and correctly) that decoherence alone does not solve the problem of the interpretation of quantum mechanics; they refer to their own Physical Collapse prejudice that the problem of quantum mechanics is to find a concrete mechanism that breaks unitarity; they misinterpret [9] as an argument to say that decoherence is the source of the breaking of unitarity of RQM; and thus conclude, mistakenly, that there is a problem. Breaking of unitarity is a trivial consequence of the perspectival aspects of RQM. That is: evolution of \emph{anything} is unitary, as long as it refers to how this ``anything'' affects something considered external to it.}

As we clearly showed in our assessment, the answer offered by Rovelli to how is it possible that a stable world, described by variables that are not relative, emerge from the quantum world, where variables are always relative, begs the question. As usual, there is no answer in Rovelli's response to this objection. And, of course, a relational theory that cannot answer such a question is in grave danger.

As for the relation between the decoherence argument in [9] and the breakdown of unitarity, in the assessment we show that, in order to explain the emergence of stable facts, it is necessary to assume stable facts beforehand, i.e., a breakdown of unitarity between the environment and the system is needed. We clearly acknowledge Rovelli's more basic explanation for the breakdown unitarity as a ``consequence of the perspectival aspects of RQM'' i.e., the story of relative acquisition of definite values by means of an interaction between systems. However, as we explain in detail in our assessment, we find such a story as vague as the collapse story in textbook quantum mechanics. We then explore whether decoherence could save the day, but show it cannot.

\section*{X. ONTOLOGY}
\hspace{\parindent} {\bf There is an interesting observation in MOS, regarding the quantum event ontology. They observe that a quantum event given by the manifestation of the position of a particle can be precisely located in spacetime, but not so for example for the momentum. This fact was noted earlier, in fact it was clear to Max Born himself [20] in his 1925 formulation of quantum theory [17, 18], and is emphasized in Niels Bohr long ruminations about complementarity between position and momentum. This is general in quantum theory: since sharp momentum implies spread position, attributing a sharp momentum to a particle is not a statement regarding a sharp spacetime location.}

{\bf I think that the impossibility to always sharply locate values of variables is just an indirect aspect of the non locality of quantum theory. RQM does not aspire to get rid of all non local aspects of quantum theory, which are just aspects of Nature. The Schrodinger equation of two particles describes physics that in a very precise sense is non-local. Nature has certain some local aspects, but we should not impose to Nature the prejudice that everything works as in relativistic \emph{classical} field theory.}

{\bf Another objection in MOS is that RQM leads to solipsism or ``quasi-solipsism'' (whatever this means). This in fact is a common objection to RQM in popular online forums, but is a funny one for a perspective entirely based on relations! It is true that RQM assumes that variables take value relative to systems, but nothing in RQM prevents the world to include complex ``observers'' like humans, that can converse among themselves and compare what they ``observe'' as observers. In fact, the theory itself guarantees that if two such observers measure (this time the term is appropriate) the same quantity and compare notes, they find agreement. This was discussed in great detail in [7] and was crucial to show that RQM is not incoherent [23, 24], as was sometimes feared in its early days.}

Here, Rovelli either fails to understand his own proposal, or is using a double standard. RQM simply cannot be used to claim that two observers comparing notes are going to actually find agreement. That would be an objective state of affairs of the world. Within RQM, we can only say that, when observer $A$ asks $B$ for his results, she is going to hear a result consistent with her own description. The converse holds for $B$. But this is clearly a different statement than to say that $A$ and $B$ agree on the values assigned. The former is, of course, very close to what happens in a solipsistic description of the world: I can explain everything I see as being how my mind works, and no observer, phenomenon, or information can make me change my mind. There is no ``agreement'' between what I see and what someone else sees, simply because there is nobody else seeing things.

{\bf There is an aspect of the ontology that is left open in the RQM literature: whether a system ontology or a quantum event ontology is more convincing. I think that RQM is compatible with both, and each has its own appeal. In the first, we assume that what exists is an electron, and its manifestations are sparse quantum event. In the second, we assume that what exists are sparse quantum events and we call electron their ensemble and their dynamical relations. I think that both views are viable. After all, we can say that there is a chair, and its manifestation are all the perspectives on it, or we can say with Hume, that a chair is nothing else that the coherent ensemble of all its manifestations. I see no reasons for which QM should decide a metaphysical issue that is open independently from quantum theory itself.}

{\bf A similar remark hold for the formulation of the theory as a principle theory in terms of information. This perspective was emphasized in [7], and has had a determined influence on the later development of diverse informational interpretations of quantum physics, including Qbism [25]. These ideas have lead to important successes in the reconstruction of quantum theory from information theoretical axioms [26–28]. But the risk of excessive emphasis on the language of information and the ambiguity of the word ``information'' which is used with wide variations of meaning risks to push towards what seems to me an excessive instrumentalist stance. If there is one thing I agree with MOS is that what we wish to get from quantum theory is a credible account of, as far as we know, the world works.}

This response assumes that RQM has two available choices for an ontology, as if both of them were clear, and the only open question was which one is more convincing. Rovelli is simply ignoring our arguments against his proposed ontology for RQM. If the idea is to define quantum events as the actualization of definite values, it must be the case that, at least the set of possible events, as well as the time at which such actualizations occur, should be well-defined (at least probabilistically). However, as we have explained above, this is not the case. The description offered is obscure, it involves information not codified in the theory and, in Rovelli's words, it ``has nothing to do with the foundation of RQM''. %Things, of course, get much worse if we follow Rovelli in regarding the wave function (even if relative) as epistemic. In that case, Rovelli would be defining his ontonolgy in epistemic terms, while any realistic theory should do just the opposite.

\section{XI. EPR}

\hspace{\parindent} {\bf The paper [29] analysis the EPR scenario at the light of RQM. MOS object to that analysis posing the following problem: The two spatially separated observers measure spin ``along the same axis, at space-like separation from each other. Suppose, moreover, that, in accordance with the possibilities allowed by RQM, they both obtain spin-up. After performing such a measurement, both of them travel to the mid-point between their labs and, at the same time, announce their results. What would then happen?'' MOS argue that according to RQM a contradiction may emerge because nothing prevents A and B from reporting the same spin, against angular momentum conservation. This is a factual mistake. What ``happens'' in RQM is how a system affect another system. Without specifying with respect to which system are the variables taking value, the question is meaningless. MOS want to avoid referring to either $A$ or $B$ as observers. This is possible in RQM. It suffice to ask how A and B affect an external system, call it $P$. The probability that $A$ and $B$ report contradictory results can be calculated using standard quantum mechanics, and is zero. The mistake of MOS is the common source of confusion in the EPR scenario, according to RQM: forgetting that A and B are themselves quantum systems and treating them classically. If we assume that they are classical, we generate the contradiction. If we keep their quantum nature in mind, there is no contradiction in any set of events relative to a single system. It is to juxtapose events relative to different systems that creates the illusion of a conflict. Facts are genuinely relative [30–33], a point of view ``from outside the universe'' is genuinely excluded [34].}

Here, Rovelli is attacking a fake opponent to RQM he himself created. We never claim to find a contradiction in the EPR scenario, so his whole response above is simply irrelevant. What we do find is that RQM is either incomplete or solipsistic, in a way that it becomes impossible to claim that the theory is empirically adequate. To see this, we are going to reproduce here the symmetrical EPR argument in very simple terms.

Given that RQM fully embraces that different observers may give different descriptions of the same sequence of events, it is natural to wonder whether RQM could lead to a situation in which different observers get different answers out of measurements of the same system. If so, wouldn't that be paradoxical? Moreover, even if shown to avoid inconsistencies, such a state of affairs would seem hard to reconcile with the fact that descriptions made by different observers are (generally) found to be consistent, i.e., with the fact that observers seem to be able to ``compare notes'' in a consistent manner.

In response to this type of worries, it is pointed out that, within RQM, the information possessed by different observers cannot be compared in the absence of physical interactions between them. To see what is meant by this, consider a Wigner's friend scenario, in which, inside a sealed lab, $O$ measures property $q$ of system $S$, with values $1$ and $2$, and $P$ stands outside. Now, according to RQM, even if $O$ finds $q=1$, it is still possible that, when $P$ later measures $S$, she finds $q=2$. Isn't that inconsistent? The point is that, within RQM, what is meaningful to ask is whether $P$'s measurement of $S$ coincides, not with what $O$ found when he measured $S$---as those measurements cannot be directly compared---but with what $O$ answers $P$ that he found when $P$ asks $O$ about his result.

The important point, it is argued, is that quantum mechanics guarantees the sort of consistency that RQM requires. This is because, after $O$ measures, $P$ assigns the $S-O$ system a superposition of $S$ measuring $q=1$ and $q=2$. Therefore, when $P$ measures $S$, she ``collapses'' or ``actualizes'' the state of the full $S-O$ system to one of those two terms, ensuring that the second measurement, i.e., asking $O$ what he got, yields an answer that is consistent with his measurement of $S$. 

In sum, RQM does not demand for measurements of the same system by different observers to always coincide---that would be an observer-independent requirement, which is meaningless within RQM. Instead, RQM only demands for the sequence of events of each observer to be self-consistent. This, it is argued, is enough to ensure the empirical success of RQM, and to remove potential charges of solipsism against it.

Before moving on, it is important to point out that, in this whole explanation, which is all Rovelli offers on the matter, the question as to what $O$ \emph{experiences} (in case O is a conscious being), in the event that he initially found $q=1$, but $P$ latter finds $q=2$, is never addressed. Does $P$'s measurement erases $O$ memory? Does $O$ remember his previous result? No answers to such questions are ever offered.

In any case, a more challenging context to test these ideas is an EPR experiment, in which a singlet is created, and one particle of the pair is sent to $A$ and the other to $B$ for they to measure them at space-like separated events. According to RQM, since $A$ and $B$ are space-like separated, there cannot exist an observer with respect to which both of these outcomes are actual. Therefore, it is meaningless to compare the results of $A$ and $B$. As a result, $A$'s and $B$'s results are fully independent and they could very well coincide, even if measured along the same axis.

However, when $B$'s particle is back into causal contact with $A$, she could go on and measure it. If so, it is argued that, as in the Wigner's friend case, quantum mechanics guarantees that, regardless of $B$'s previous result, $A$ will find the opposite of what $B$ found on his particle. Moreover, if $A$ goes on and asks $B$ what he found, regardless of what $B$ found before, $B$'s answer will be consistent with what $A$ found when she measured $B$'s particle. Finally, it is argued that, if a third observer asks both $A$ and $B$ for their results, perfect anti-correlation is guaranteed in the answers. The arguments for all this are analogous to the one discussed above regarding the Wigner's friend scenario: a measurement of $B$'s particle by $A$ collapses to a branch that guarantees the required correlations, and similarly for the third observer.

The lesson derived from all this is that, what RQM provides, is not the collection of all properties relative to all systems (such collection is assumed not to make sense), but one self-consistent description per system. Yet, systems can interact, and any system can be observed by another system. This, implies that any two observers can in fact communicate and, moreover, that either account of such interaction is guaranteed to be self-consistent. Inconsistencies arise only if one insists on an absolute state of affairs, obtained by juxtaposing descriptions relative to different observers. It is concluded that RQM is the stipulation that this self-consistent, individual descriptions, is all one can talk about and, more importantly, that this is enough ``for describing nature and our own possibility of exchanging information about nature (hence circumventing solipsism)''. We, on the other hand, are rather skeptical about these claims, as we should see.

The point is that, in all scenarios considered above, there is an asymmetry between the observers involved, (e.g., $P$ is outside and measures after $O$; $A$ goes and measures $B$'s particle, and not the other way around), which is exploited to (allegedly) avoid trouble. However, when such an asymmetry is removed, the problems with the proposal become clear. To see this, consider again an EPR scenario, and suppose that an experiment is conducted as follows:
\begin{enumerate}
\item $A$ and $B$ receive a particle each and they measure it along the same axis, at space-like separation from each other; suppose that, in accordance with the possibilities allowed by RQM, they both obtain spin-up. 
\item After performing such a measurement, both of them travel to the mid-point between their labs and, at the same time, announce their results.
\end{enumerate}

What would happen then? There's a list of alternatives:
\begin{itemize}
\item One possibility is that, contrary to what RQM asserts, it is in fact metaphysically impossible for them to find the same result. This, of course, would mean that RQM is incomplete, which would be devastating for the proposal. 
\item Another option is that, for some reason, either $A$ or $B$ would announce the wrong result. This is exactly what is argued to happen in the asymmetric scenarios considered above, such as in Wigner's friend scenario, if $P$ asks $O$ about his result, in a case in which $O$ and $P$ find different values for $q$. However, in such a scenario, when $P$ measures $S$, she must interact with the whole $S-O$ system, so it is at least plausible that such an interaction could modify the state of $O$ in such a way that he ends up believing that he obtained a result for $q$ different from what he actually did (as we said above, Rovelli is mute on this issue). In the symmetric EPR scenario under consideration, on the other hand, an explanation of that sort seems out of place. To begin with, the situation is fully symmetric between $A$ and $B$, so it is not clear how it could be decided who would end up announcing the wrong result; and certainly, RQM does not have the means to do so. Moreover, unlike the Wigner's friend case, there does not seem to be any sort of mechanism that could explain why would, either $A$ or $B$, report a different result from what they actually obtained. Only stipulating that this is what would need to happen is, of course, not satisfactory.
\item Yet another option would be for both $A$ and $B$ to report either the right or the wrong result. That, of course, would preserve the symmetry, but would imply both reporting the same result, which would be empirically inadequate.
\item A final way out would be to allow for a total disconnect either between what each observer announce and what they think they announce, or between what they announce and the other hears. But that would clearly lead us deep into a solipsistic description.
\end{itemize}

From all this, it is clear that RQM has no sensible way to compare different descriptions of the same sequence of events, something we expect from any reasonable scientific theory. As we have said, RQM only establishes that an observer's view will be consistent with itself, but fully fails to establish any sort of relation between the descriptions provided by different observers. The specter of solipsism endures.

\section*{XII. LOCALITY}

\hspace{\parindent} {\bf The discussion of locality in MOS reflects the general pattern. MOS take locality as the requirement that there are local beable in the sense of Bell that obey Bell local causality. They then argue that his is not the case in RQM. This is true in fact, and is precisely what is argued in the paper [35]. What is then argued in [35] that in the light of RQM this non locality is a simple consequence of the fact that some variables are not defined at some times. MOS equivocate on this statement, reinterpreting it as meaning that the variables are well defined but not deterministically determined. No surprise on this, because MOS assume that there is always a unique determined state, this is part of their assumptions, assumption abandoned in RQM. Hence they misunderstand [35] and wrongly conclude that the conclusion of [35] is ``false''.}

What is actually claimed in [35] is that ``the failure of `locality' in the sense of Bell, once interpreted in the relational framework, reduces to the existence of a common cause in an indeterministic context'', and that Bell's definition may not be appropriate in this case because it ``does not capture finely enough the intuitive idea of an indeterministic `no superluminal causal influence' as Bell would have liked''. Now, Rovelli changes course and claims that Bell's definition of locality does not apply in the context of RQM because ``some variables are not defined at some times''. This change in position, does not help his cause because, for an observer to the future of $A$ and $B$, it is simply not the case that ``some variables are not defined at some times''. For such an observer, according to [35], all relevant variables in his past light cone are, in fact, defined. Therefore, at least for him, there is no escape to Bell's conclusion that the correlations between $A$'s and $B$'s result cannot be explained by the postulation of a local common cause, be it deterministic or indeterministic. We reiterate that the conclusion in [35] is wrong.

\section*{XIII. CONCLUSION}

\hspace{\parindent} {\bf Altogether, the paper [2] turns out to be an exercise in misunderstanding. This is not surprising, because the paper is really an attempt to make sense of RQM in the context of the hypotheses underlying a Physical Collapse interpretation of Quantum Theory. It is an attempt of interpreting RQM on the basis of assumptions rejected by RQM.}

{\bf The article could be read as an indirect way to promote the Physical Collapse perspective by using RQM as a straw man.}

As we have show throughout this response, there is not a shred of evidence to support this accusation. In the assessment, we make use of nothing, except the author's own claims and statements, some logical combinations of those, and the analysis of specific cases that allow for a detailed analysis of what RQM really is able to accomplish.

{\bf It is written as if it was based on an ``objective'' position, based on commonly accepted assumptions. But the facts and assumptions that it gives for granted are far from being commonly accepted. They are the specific assumptions of the authors' ``camp''. The authors have employed this strategy elsewhere [36].}
 
The author is, in a sense, right. We have used before the strategy of bringing to light misleading arguments, unjustified claims and inconsistent proposals, which have acquired certain popularity within the community. We do so in an attempt to revert the confusion they are creating and to serve as a warning to younger generations about important problems that are (purposely or mistakenly) being ignored. We feel that such is the duty of the scientific community and, in particular, of the scientists that examine those proposals in detail. We think it is sometimes the case that proponents of an idea so fall in love with it that they fail to see its shortcomings. Other times, they might be well aware of the problems, but choose to ignore or hide them. Here, we do not follow Rovelli's lead, and refrain from speculating which is the case in the present instance.

Regarding [36], the authors of that work attempt to account for the emergence of classicality in cosmology in terms of the purely unitary part of the quantum evolution. Rovelli, on the other hand, emphasizes in this response to our assessment that the purely unitary part of the quantum evolution cannot be everything. So, the question for Rovelli is now, with which of our assessments does he disagree with and why? Moreover, in order to defend his credibility, he should clearly explain where exactly, in our analysis of say [36], we rely on assumptions that are particular to what he calls ``our camp''. Once more, these vague accusations are not accompanied by specifics.

{\bf This invalidates the value of the assessment.}

It is true that, some of us, have worked, among many other things, on collapse models. However, trying to dismiss our careful evaluation of RQM, without engaging with our actual arguments, and simply by pointing that out, gets pretty close to an \emph{ad hominem} attack.\footnote{Following Rovelli's logic, should we, in turn, infer that, for instance, when Rovelli stated: ``The string planet is infinitely less arrogant than ten years ago, especially after the bitter disappointment of the non-appearance of supersymmetric particles'' or ``It is possible that the two theories could be parts of a common solution … but I myself think it is unlikely. String theory seems to me to have failed to deliver what it had promised in the '80s, and is one of the many `nice-idea-but-nature-is-not-like-that' that dot the history of science. I do not really understand how can people still have hope in it.'' (Interview used in the article ``String Theory Meets Loop Quantum Gravity'' by Sabine Hossenfelder, Quanta Magazzine, June 16, 2016 issue.), he was doing so, not out of his sincere assessment of String Theory and its accomplishments, but just as a deceitful mean to promote his Loop Quantum Gravity ``camp''?}. If Rovelli cannot point out the mistakes he claims we have made in criticizing his proposals, his claim that our assessment of RQM is invalid, is clearly exhibited as fully hollow.

{\bf Unfortunately, this happens repeatedly in debates about the interpretation of quantum mechanics. Defenders of one interpretation construct detailed arguments trying to show that another interpretation is ``wrong'', failing to see that what they are doing is simply applying their own hypotheses to the logic of authors that have alternative hypotheses.}

As we have shown explicitly, and in excruciating detail, the attempt to discredit our assessment of RQM, and the defense of RQM in his response, are simply a series of misrepresentations and linguistic acrobatics that, if anything, help to further exhibit the serious shortcomings of Rovelli's proposal.

{\bf The difficulty of quantum mechanics is not about carefully articulating consequences of a shared conceptual framework: it is to understand which conceptual framework is more promising for making sense of the theory. As mentioned in the introduction, every interpretation of QM includes conceptual steps hard to digest. The exercise of criticizing the details of an interpretation without accepting these conceptual steps is futile. I am sure we can do better, in articulating a fruitful conversation between different ways of making sense of quantum theory.}

Here, we agree wholeheartedly with Rovelli. We only find it regrettable that, in assessing his own proposal, he departs so dramatically from his own recommendations.

\section*{References}

\begin{description}
\item[] [1] Federico Laudisa and Carlo Rovelli, ``Relational Quantum Mechanics,'' The Stanford Encyclopedia of Philosophy (2021).

\item[] [2] R. Muciño, E. Okon, and D. Sudarsky, ``Assessing Relational Quantum Mechanics,'' arXiv:2105.13338. 

\item[] [3] E. Okon and D. Sudarsky, ``Benefits of Objective Collapse Models for Cosmology and Quantum Gravity,'' Foundations of Physics 44, 114-143 (2014).

\item[] [4] E. Okon and D. Sudarsky, ``The Black Hole Information Paradox and the Collapse of the Wave Function,'' Foundations of Physics 45, 461-470 (2015).

\item[] [5] D. Bedingham, S. K. Modak, and D. Sudarsky, ``Relativistic collapse dynamics and black hole information loss,'' Phys. Rev. D 94, 045009 (2016).

\item[] [6] D. Sudarsky, ``Spontaneous Collapse Theories and Cosmology,'' in Do Wave Functions jump? (Springer, 2021).

\item[] [7] Carlo Rovelli, ``Relational Quantum Mechanics,'' Int. J. Theor. Phys. 35, 1637 (1996), arXiv:9609002 [quant-ph].

\item[] [8] Carlo Rovelli, ```Space is blue and birds fly through it','' Philosophical Transactions of the Royal Society A: Mathematical, Physical and Engineering Sciences 376 (2017),arXiv:1712.02894.

\item[] [9] Andrea Di Biagio and Carlo Rovelli, ``Stable Facts, Relative Facts,'' Foundations of Physics 51, 30 (2021),arXiv:2006.15543.

\item[] [10] Claudio Calosi and Christian Mariani, ``Quantum Relational Indeterminacy,'' Studies in History and Philosophy of Modern Physics (2020).

\item[] [11] Mauro Dorato, ``Bohr meets Rovelli: a dispositional and Humanism'' (Cambridge Univeristity Press, Cambridge, 1996).

\item[] [12] Simon Kochen, ``A new interpretation of quantum mechanics,'' in Symposium on the Foundations of Modern Physics, edited by P. Mittelstaedt and P. Lahti (World Scientific, Singapore, 1985) pp. 151-169.

\item[] [13] D. Dieks, ``The formalism of quantum theory: an objective description of reality?'' Annalen der Physik 7, 174-190 (1988).

\item[] [14] D. Dieks, ``Modal interpretation of quantum mechanics, measurements, and macroscopic behaviour,'' Physical Review A 49, 2290-2300 (1994).

\item[] [15] G. C. Ghirardi, A. Rimini, and T. Weber, ``Unified dynamics for microscopic and macroscopic systems,'' Physical Review D 34, 470-491 (1986).

\item[] [16] Roger Penrose, ``On gravity's role in quantum state reduction,'' General Relativity and Gravitation 28, 581-600 (1996).

\item[] [17] M Born and P Jordan, ``Zur Quantenmechanik,'' Zeitschrift für Physik 34, 858-888 (1925).

\item[] [18] M Born, P Jordan, and W Heisenberg, ``Zur Quanten- mechanik II,'' Zeitschrift für Physik 35, 557-615 (1926).

\item[] [19] William A. Fedak and Jeffrey J. Prentis, ``The 1925 Born and Jordan paper `On quantum mechanics','' American Journal of Physics 77, 128 (2009).

\item[] [20] Herbert Capellmann, ``Space-Time in Quantum Theory,'' Foundations of Physics (2021), arXiv:2006.00503.

\item[] [21] Erwin Schrödinger, Nature and the Greeks and Science.

\item[] [22] Carlo Rovelli, ``Incerto Tempore, Incertisque Loci: Can We Compute the Exact Time at Which a Quantum Measurement Happens?'' Foundations of Physics 28, 1031-1043 (1998), arXiv:9802020 [quant-ph].

\item[] [23] Bas C van Fraassen, ``Rovelli's World,'' Foundations of Physics 40, 390-417 (2010).

\item[] [24] Michel Bitbol, ``Physical Relations or Functional Relations? A non-metaphysical construal of Rovelli's Relational Quantum Mechanics,'' (2007).

\item[] [25] Christopher A. Fuchs, ``Quantum Foundations in the Light of Quantum Information,'' in Decoherence and Its Implications in Quantum Computation and Information Transfer: Proceedings of the NATO Advanced Research Workshop, edited by A. Gonis and P. E. A. Turchi (Amsterdam, 2001) ios press ed., arXiv:0106166 [quant-ph].

\item[] [26] Philipp Andres Höhn, ``Toolbox for reconstructing quantum theory from rules on information acquisition,'' Quantum 1 (2017), arXiv:1412.8323.

\item[] [27] Philipp Andres Höhn and Christopher S.P. Wever, ``Quantum theory from questions,'' Physical Review A 95 (2017), arXiv:1511.01130.

\item[] [28] Philipp Andres Höhn and Christopher S.P. Wever, ``Quantum theory from questions,'' Physical Review A 95, 012102 (2017), arXiv:1511.01130.ist account of the quantum state, Quantum Studies: Mathematics and Foundations 7, 233–245 (2020), arXiv:2001.08626. ph].

\item[] [30] Daniela Frauchiger and Renato Renner, ``Quantum theory cannot consistently describe the use of itself,'' Nature Communications 9 (2018), arXiv:1604.07422.

\item[] [31] Kok-WeiBong,AnıbalUtreras-Alarcón,FarzadGhafari, Yeong-Cherng Liang, Nora Tischler, Eric G. Cavalcanti, Geoff J. Pryde, and Howard M. Wiseman, ``Testing the reality of Wigner's friend's observations, (2019), arXiv:1907.05607.

\item[] [32] Caslav Brukner, ``A No-Go theorem for observer- independent facts,'' Entropy 20 (2018).

\item[] [33] Marijn Waaijer and Jan van Neerven, ``Relational analysis of the Frauchiger–Renner paradox and interaction- free detection of records from the past,'' (2019), arXiv:1902.07139.

\item[] [34] Mauro Dorato, ``Rovelli's relational quantum mechanics, monism and quantum becoming,'' in The Metaphysics of Relations, edited by A Marmodoro and A Yates (Oxford University Press, 2016) pp. 290–324, arXiv:1309.0132.

\item[] [35] Pierre Martin-Dussaud, Carlo Rovelli, and Federico Zalamea, ``The Notion of Locality in Relational Quantum Mechanics,'' Foundations of Physics 49, 96-106 (2019), arXiv:1806.08150.

\item[] [36] Javier Berjon, Elias Okon, and Daniel Sudarsky, ``Critical review of prevailing explanations for the emergence of classicality in cosmology,'' (2020), arXiv:2009.09999.

\end{description}
%%%%%%%%%%%%%%%%%%%%%%%%%%%%%%%%%%%%%%%%%%%%%%%%%%%%%%%%%%
\section*{Acknowledgments}
%%%%%%%%%%%%%%%%%%%%%%%%%%%%%%%%%%%%%%%%%%%%%%%%%%%%%%%%%%
%%%%%%%%%%%%%%%%%%%%%%%%%%%%%%%%%%%%%%%%%%%%%%%%%%%%%%%%%%
 We thank the author for sending us his reply to our assessment the day before uploading it into the arXiv.
D.S. acknowledges partial financial support from PAPIIT-UNAM grant IG100120; CONACYT grant 140630; the Foundational Questions Institute (Grant No. FQXi-MGB-1928); the Fetzer Franklin Fund, a donor advised by the Silicon Valley Community Foundation. E.O. acknowledges support from UNAM-PAPIIT grant IN102219 and CONACYT grant 140630.

%%%%%%%%%%%%%%%%%%%%%%%%%%%%%%%%%%%%%%%%%%%%%%%%%%%%%%%%%%
%%%%%%%%%%%%%%%%%%%%%%%%%%%%%%%%%%%%%%%%%%%%%%%%%%%%%%%%%%
%\bibliographystyle{plain}
\bibliographystyle{apalike}
\bibliography{bibRRQM.bib}
%%%%%%%%%%%%%%%%%%%%%%%%%%%%%%%%%%%%%%%%%%%%%%%%%%%%%%%%%%
%%%%%%%%%%%%%%%%%%%%%%%%%%%%%%%%%%%%%%%%%%%%%%%%%%%%%%%%%% 

\end{document}